\documentclass[usegraphicx]{mn2e}
\usepackage{subfigure}
\usepackage{epsf}
\usepackage{amsmath}

\title[Doppler disc tomography and low mass AGN spin]
{Doppler disc tomography applied to low mass AGN spin}

\author[M.Middleton \& A. Ingram]
{Matthew J. Middleton$^{1}$ \& Adam R. Ingram$^{2}$\\
\\
1. Institute of Astronomy, Madingly Road, Cambridge, CB3 0HA, UK\\
2. Astronomical Institute Anton Pannekoek, Science Park 904,1098 XH, Amsterdam, Netherlands
}

\pagerange{\pageref{firstpage}--\pageref{lastpage}} \pubyear{2014}
\long\def\symbolfootnote[#1]#2{\begingroup\def\thefootnote{\fnsymbol{footnote}}\footnote[#1]{#2}\endgroup} 
\def\ga{\mathrel{\hbox{\rlap{\hbox{\lower4pt\hbox{$\sim$}}}{\raise2pt\hbox{$>$}}
}}}

\begin{document}

\topmargin = -0.5cm

\maketitle

\label{firstpage}

\begin{abstract}

Doppler tomography can provide a powerful means of determining black hole spin when our view to the central regions are revealed and obscured by optically thick orbiting material, and can provide an independent estimate that does not suffer as many degeneracies as traditional methods. For low mass AGN, time-dependent obscuration is expected to leave a signature in the changing spectrum of the disc emission which extends into the soft X-ray bandpass. We create a spectral model incorporating Doppler tomography and apply it to the case of the low mass (8$\times$10$^{5}$ M$_{\odot}$) AGN,  RX J1301.9+2747 which shows unusual timing properties in the form of short-lived flares that we argue are best explained by the orbit of a window through an optically thick wind. Modelling the phase-resolved spectrum over the course of the highest data quality flare indicates a very low spin even when we relax our constraints. This is the lowest mass AGN for which a spin has been measured and the first via this technique. We note that, as the mass and spin are very low, this appears to favour supermassive black hole (SMBH) growth by chaotic rather than constant accretion. 

\end{abstract}

\section{Introduction}

The theory of general relativity predicts that astrophysical black holes can be fully defined by only their mass and angular momentum (the `no-hair' theorem), the latter represented by the dimensionless quantity of `spin', $a_{*}$ (defined as $Jc/GM_{\rm BH}^{2}$). which may take values between -0.998 for maximal retrograde and 0.998 for maximal prograde spin relative to the orientation of the accretion flow. Both the mass and spin are initially set by the formation mechanism, although on very long timescales accreted material may add to both, with the change in spin dependent on the orbital direction of the infalling gas. In black hole binary systems (BHBs) the supernova explosion can expel angular momentum of the progenitor star leading to low natal spins that increase over time through accretion (Bardeen 1970). In active galactic nuclei (AGN), however, the origin of the spin of the SMBH is unclear as the growth mechanism could be via continuous accretion of gas leading to high spins, or chaotic (e.g. Nayakshin, Power \& King 2012) with a distribution of infall angles and lower spins (Volonteri et al. 2005). These two scenarios yield different predictions for the spin distribution of low mass SMBHs  (10$^{5-6}$ M$_{\odot}$) at low redshift (Fanidakis et al. 2011). In order to determine the growth mechanism it is therefore important to find a reliable manner in which to measure the spin of SMBHs at the low end of the mass scale.

The process of accretion liberates the gravitational energy of infalling material, efficiently converting it to thermal energy in a viscous accretion disc (where the viscous liberation of angular momentum is mediated by the MRI: Balbus \& Hawley 1998). The emission from such a geometrically `thin', optically thick disc peaks just outside of the innermost stable circular orbit (ISCO) where material can safely reside in a Keplerian orbit before taking a laminar plunge to the event horizon. The black hole spin sets the position of the ISCO: at low spins ($a_{*}$ = 0) the orbit rests at 6~$R_{\rm g}$ ($R_{\rm g}$ defined as $GM/c^2$) but with increasing prograde spin (up to $a_{*}$ = 0.998) the orbit moves inwards through frame dragging to 1.24~$R_{\rm g}$. It is also feasible for the spin to be retrograde, i.e. counter-aligned to the rotational direction of the accreting gas, where the ISCO moves out to 9~$R_{\rm g}$ (for $a_{*}$ = -0.998). The change in location of the ISCO leads to two directly observable effects: the rest-frame emission from the accretion disc is considerably hotter for a higher spin and distortion of the emission due to the relative motion of the approaching and receding sides of the disc (Doppler broadening) and relativistic effects (gravitational redshift) increases. 

Observationally, the energy spectrum of accreting black holes is only dominated by disc emission at higher rates of accretion (Remillard \& McClintock 2006). In the specific case of BHBs, the inner disc temperature peaks in the X-ray band allowing estimates of the spin from the spectrum (e.g. Steiner et al. 2011, McClintock et al. 2006; Middleton et al. 2006; 2014) using models incorporating relativistic effects and radiative transfer (e.g. Davis et al. 2005). At lower rates, a large fraction (sometimes the dominant component) of the emission originates in a Compton up-scattered thermal/non-thermal tail. The re-illumination of the disc by the Compton tail leads to a characteristic reflection spectrum (Fabian et al. 1989) seen ubiquitously in both BHBs and AGN, with strong iron fluorescence features (notably Fe K$_{\alpha}$ but also Fe L$_{\alpha}$, e.g. Fabian et al. 2009). Under the assumption that the inner disc edge extends down to the ISCO and that the orbits are Keplerian, the spin can then be estimated from the amount of smearing in these lines (due to Doppler broadening, e.g. Brenneman \& Reynolds 2006).

Both of these methods have been widely applied in attempts to probe the spin of stellar mass BHs in BHBs, although disagreements do occur (Reis et al. 2008; Kolehmainen \& Done 2010). In high mass AGN, the Fe K$_{\alpha}$ line is widely observed, leading to measures of near maximal spin for those with reflection dominated spectra (Reynolds 2013). Although such spectra may be complicated by absorption, the shape of the high energy Compton reflection hump onto which the Fe K$_{\alpha}$ line is superimposed aids in distinguishing between models (Risaliti et al 2013),  with the highest quality data allowing the broad and narrow components of the Fe K$_{\alpha}$ line to be unambiguously separated via variability properties (Zoghbi et al. 2012). Applying this method to low mass AGN is problematic as they are selectively detected at higher mass accretion rates (Greene \& Ho 2004) where, analogous to BHB spectra the Compton tail of emission is often considerably weaker (Jin et al. 2012) and winds, driven from the disc, will be present and may distort our view of the emission. 

Modelling the disc emission directly is non-trivial in AGN as much of the emission is out of the X-ray bandpass and is heavily attenuated by the UV opaque interstellar medium. In addition, whilst modelling changing disc spectra over time is possible for BHBs - thereby ensuring a fixed position for the ISCO (see Gierli{\'n}ski \& Done 2004) - this presents a significant obstacle for AGN where accretion driven changes occur on timescales of decades to millennia.  New models for low mass AGN ($<$ 10$^{7}$ M$_{\odot}$) predict that the disc emission should extend into the soft X-ray bandpass (Done et al. 2012) and a subsequent attempt to measure the spin from continuum fitting in a single epoch has yielded $a_{*} <$ 0.86 for a ~10$^{7}$ M$_{\odot}$ AGN, in apparent conflict with reflection modelling in the same source (Done et al. 2013) and reinforcing the issues associated with traditional methods for measuring the spin. Although this result cannot constrain the SMBH growth mechanism, a reliable measurement of low spin in a local, very low mass AGN would, in principle, strongly favour a chaotic growth scenario (Fanidakis et al. 2011). 

\begin{figure}
\center
\includegraphics[width=3in, angle=0]{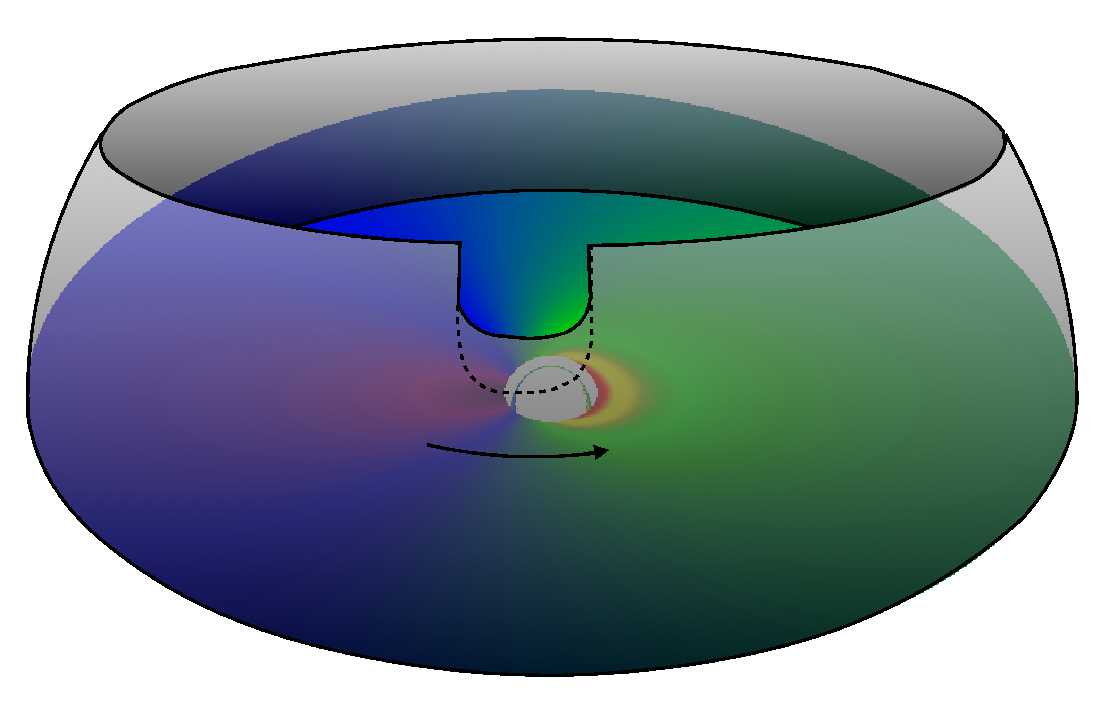}
\caption{Physical schematic of the model. The disc emission (with colours indicating blue-shift and red-shift respectively) including the full ensemble of relativistic effects (as determined from {\sc geokerr}: Dexter \& Agol 2009) and dependent on spin and inclination) and window size (dependent on the flare duration and shape) are taken from the best fit to the phase-binned data (see Table 2 for details). The Keplerian rotation of the wind leads to the embedded window - seen as an irregularity in the edge of the wind - passing over the approaching, blue-shifted side of the disc and then the receding red-shifted side of the disc. The emission when the window is not visible comes from the disc up to the wind radius with some small component potentially scattered from inside the cone of the wind (though the majority of flux from inside the wind radius is down-scattered to lower energies, e.g. Titarchuk \& Shrader 2005). The dotted line indicates the 90\% maximum inclination of the window (see Table 2).}
\end{figure}

\section{Doppler Tomography}

The now accepted prevalence of dense outflows from high accretion rate AGN (Tombesi et al. 2010) allows us to apply a new technique to these sources: Doppler tomography. Traditionally used to map the accretion flow about white dwarfs using the optical line emission (e.g. Marsh \& Horne 1988), here instead we can use Doppler Tomography to measure the black-hole spin. Such winds are expected to be inhomogeneous along their surfaces due to hydrodynamic instabilities (e.g. Proga \& Kallman 2004; Takeuchi et al. 2014). As a consequence, should our view favourably align with the edge of the wind then this irregular structure will lead to the inner disc being obscured and then revealed on the orbital timescale (as any launched material will still possess orbital angular momentum, Figure 1). 

Such obscuration has been considered by Risaliti et al. (2011) in the case of high mass AGN, where the shape of the Fe K$_{\alpha}$ line  - due to reflection from optically thick material in the disc (Fabian et al. 1989) - changes in response to the portion of the disc revealed by the structure of the wind (see also Ingram \& Done 2012). Conversely, for low mass AGN, where the disc extends into the soft X-ray band, it should be possible to observe and model the emission from the approaching and receding sides of the disc {\it directly}. 

In order to create a model to describe such a physical scenario, we utilise a freely available code to trace photon paths from an accretion disc, ({\sc geokerr}: Dexter \& Agol 2009) and account for the ensemble of relativistic effects (e.g. gravitational redshift, relativistic aberration; see Appendix). By obscuring all emission other than an orbiting circular aperture (the shape has little effect as discussed in the Appendix) that moves across the face of the disc inclined to the observer, we obtain tight predictions for the changing spectrum where we observe first the Doppler blue-shifted and then red-shifted disc emission (see Figure 1). The predictions depend on the size, relative latitude of the window and how fast this moves (see Appendix). The maximum rest-frame disc temperature is determined according to standard theory (Shakura Sunyaev 1973) from the BH mass, accretion rate and spin, and is shifted to account for Compton scattering in the photosphere of the UV bright disc (Done et al. 2012). As the observed spectrum as a function of time is strongly dependent on the spin and inclination, we should be able to obtain tight constraints even for modest quality data. This is then an extremely powerful, yet arguably simple technique, where we can directly fit the model to the data in several time/phase bins and constrain the spin in a fully self-consistent way. 

As the nature of the disc within such a wind is unknown we also build into our model a means to account for differences in the emissivity profile as a result of advection of radiation in a large scale-height, radiation pressure dominated disc (see Abramowicz et al. 1988) or the colour temperature correction (which should depend on the amount of turbulence).

\section{RX J1301.9+2747}

RX J1301.9+2747 is located in the Coma cluster and optically identified (Sun et al. 2013) as a very low mass AGN at 8$\times$10$^{5}$ ($\pm$0.5 dex) M$_{\odot}$. The source shows an almost entirely unabsorbed disc-dominated spectrum in the soft X-ray bandpass (with a very weak power-law tail, analogous to the high-soft state of BHBs: Remillard \& McClintock 2006) seen in data collected by both {\it Chandra} and {\it XMM-Newton} (Sun et al. 2013). The nuclear point source shows an unexplained and large (around an order of magnitude) optical excess above the extrapolated best-fitting disc model to the quiescent emission (Sun et al. 2013). Archival data extending back to 1990 (Dewangan et al. 2000) show that, from the earliest observations (with {\it ROSAT}) of the source, the lightcurve is remarkably flat and featureless, with short flares of $\sim$2~ks duration (see Figure 2), observed at a spacing of $\ge$ 18~ks (Sun et al. 2013). This is hitherto unique behaviour for an AGN, which generally show a range of variability properties on a range of long timescales (Gonzalez-Martin \& Vaughan 2012). 

\begin{figure}
\begin{center}
\begin{tabular}{l}
 \epsfxsize=8cm \epsfbox{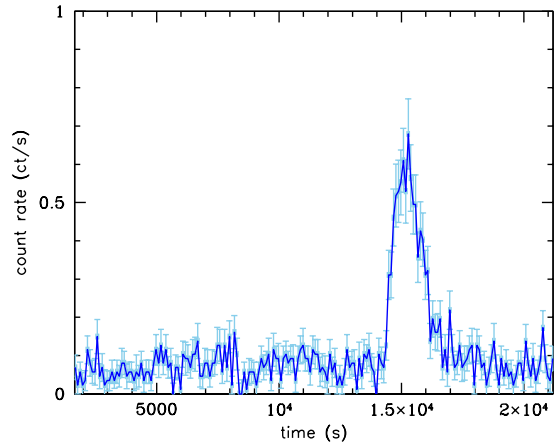}
\end{tabular}
\end{center}
\caption{Background subtracted (100s binned) PN lightcurve for RX J1301.9+2747 over the 0.2-10~keV range. Such short-lived flares, seen in both {\it Chandra} and archival {\it ROSAT} data are the sole notable feature and represent hitherto unique behaviour for an AGN.}
\end{figure}

The highest quality X-ray data for RX J1301.9+2747 was obtained by {\it XMM-Newton} in December 2000 with the X-ray spectrum of the characteristic `flare' (Figure 2) showing a strengthening and broadening of the disc component relative to the non-flaring spectrum, also seen in the spectrum of the {\it Chandra} data (Sun et al. 2013). Given the association with the disc component rather than the Compton tail this is highly unlikely to be due to reconnection (where we would expect a far greater increase in the power-law tail) and is distinctly unlike the flares seen at very low rates of accretion onto Sgr A* (e.g. Degenaar et al. 2013 and references therein). Due to the rise timescale of $<$ 1ks, such a change is unlikely to be due to mass accretion rate driven variations as these operate on viscous timescales (Shakura \& Sunyaev 1973) given by:

\begin{equation}
t_{visc} (R)\approx 10^{7} s \left(\frac{M_{5}}{8}\right)\left(\frac{\alpha}{0.1}\right)^{-1}\left(\frac{H/R}{0.01}\right)^{-2}\left(\frac{R}{6R_{g}}\right)^{3/2}
\end{equation}

where $\alpha$ is the dimensionless viscosity parameter (Shakura \& Sunyaev 1973), $M_5$ is the BH mass in units of 10$^5$ M$_{\odot}$ and H/R is the scale height of the disc. Whilst we do not expect to see variability originating from only a single radius on a single viscous timescale (as variability propagates radially: Lyubarskii 1997) it is instructive to see that, for reasonable values of the parameters in the above equation, the viscous timescale at 6~$R_{\rm g}$ is $\sim$10$^{7}$~s for RX J1301.9+2747, which is far longer than the flare rise timescale of $<$1~ks. If the BH is maximally spinning, the disc may extend down to an inner edge at $\sim$1$~R_{\rm g}$, where the viscous timescale will be $\sim$7$\times$10$^{5}$~s: still far longer than the flare rise time. If the disc is instead geometrically thick, we could push H/R to 0.1 and still find a viscous timescale at 1~$R_{\rm g}$ of 7~ks $>$ 1ks.

We must also consider whether the timescale of the flares could be a signature of a thermal process (i.e. the timescale for emission at some radius in the disc): $t_{\rm ther} = t_{\rm dyn}/\alpha$ where $t_{\rm dyn} = 1/\Omega_{\rm k}$ and $\Omega_{\rm k}$ is the Keplerian angular velocity. Assuming $\alpha$ = 0.1, then we would need the energy deposition to occur at 8.6~$R_{\rm g}$, i.e. very close to the ISCO. Whilst this is not unfeasible (the emission may well be dominated by the most inner regions), to make the release of thermal energy a flaring process would require a thermal instability (e.g. Shakura \& Sunyaev 1976). Those sources which are radiation pressure dominated are most prone to such instabilities, implying that such flaring should be commonly seen in more massive AGN; yet the flaring behaviour of RX J1301.9+2747 is clearly extremely rare, implying that the flaring is not some general thermal timescale behaviour of the disc. We can also compare with BHBs to show that this flare is very rapid compared with standard accretion driven changes. The soft state in BHBs is characterised by a disc dominated spectrum (Remillard \& McClintock 2006) and very little variability, in direct analogy to RX J1301.9+2747 during the stable quiescent intervals. In the hard state, we see strong variability on timescales longer than 0.1~s, although the hard spectrum bares no resemblance to the spectrum of RX J1301.9+2747. Scaling from a ~10~M$_{\odot}$ to a ~10$^{6}$~M$_{\odot}$ BH implies a hard state analogy with the mass of RX J1301.9+2747 could show variability on timescales longer than ~10~ks. So, even considering this system to vary like a BHB in the hard state yields a characteristic variability timescale an order of a magnitude too large. A more extreme and perhaps more salient example may be the rare ms flares seen in the soft state of Cyg X-1
(Gierlinski \& Zdziarski, 2003), which, based on mass scaling would give a similar timescale to the flares in  RX J1301.9+2747. However, unlike the flares in RX J1301.9+2747, the ms flares are imprinted onto variability on a range of timescales. RX J1301.9+2747 therefore displays completely different behaviour to accretion rate driven variability routinely observed in BHBs and other AGN, which vary stochastically across a range of timescales and never show isolated flares either side of constant quiescent flux (Gonzalez-Martin \& Vaughan 2012). We thus conclude that a dynamical origin is a far more plausible interpretation.

A natural explanation for both the short timescale flares and the optical excess is a changing line-of-sight to the inner regions on orbital timescales, i.e. `windows' through a highly optically thick wind which otherwise restricts our view to the radius at which it is launched. As we have mentioned, in practice we would not expect a hole through the wind but rather an inhomogeneous wind edge and fortuitous sightlines (see Figure 1). Such an optically thick shroud would leave little imprint of absorption in the X-ray spectrum but would re-process a portion of the soft X-ray emission to lower energies via Compton down-scattering in the cooler wind (Titarchuk \& Shrader 2005), as inferred for the BHB SS433 (Dolan et al. 1997). Whilst potentially large amounts of the soft flux may be scattered out of the wind-cone (King 2009), the remaining emission passing through the wind will then lead to excess optical emission as seen in RX J1301.9+2747, supporting our physical picture. As a corollary, the presence of such a wind implies the accretion rate in RX J1301.9+2747 is likely to be at or above Eddington (Shakura \& Sunyaev 1973; Poutanen et al. 2007; King 2009). Although the clear presence of the accretion disc in the X-ray bandpass indicates high fractions of Eddington, the true accretion rate is likely to be far greater than derived from the SED fitting (i.e. $>$ 0.1 Eddington: Sun et al. 2013) due to reprocessing and scattering of flux out of the cone of the wind. 

\section{Data extraction and spectral fitting}

{\it XMM-Newton} observed RX J1301.9+2747 on 10/12/2000 for a total exposure time of 24 ks (PN) and 28 ks (MOS) respectively with the source 7 arcminutes off axis (resulting in a reduction in data quality due to vignetting). We reduced the raw event mode data using SAS version 13.0 and up-to-date calibration files, following standard extraction procedure\footnote{http://xmm.esac.esa.int/external/xmm\_user\_support\\/documentation/sas\_usg/USG/}.

We extracted the high-energy (10-15 keV) full field lightcurve, creating good time intervals free of any dominant soft proton flares. From images created using {\sc xselect} we defined source and background regions (45 arcsec in radius) on the same chip (with the background region selected to avoid read-out direction) and extracted lightcurves and spectra (source, background and response files) filtering for standard patterns and flags. We group the latter (using {\sc grppha}) to have a minimum of 20 counts/bin thereby ensuring Gaussian statistics for chi-squared fitting.

The lightcurve of the MOS cameras (shown in Sun et al. 2013) is of a longer duration than the PN (due to a longer exposure although through-put is far lower) and shows the tail end of a `flare' at the start of the observation. This allows us to place a lower limit on the distance from the BH to the window if we were to assume the limiting case of a periodic feature (which we cannot at present determine from {\it ROSAT} data due to the observing gaps). Assuming a Keplerian period of $\sim$18~ks this would correspond to a distance of $\sim$80 $R_{\rm g}$ for an 8$\times$10$^{5}$ M$_{\odot}$ BH.

\begin{table}
\begin{center}
\begin{minipage}{45mm}
\bigskip
\caption{Phase bin information}
\begin{tabular}{lc|c|}
  \hline

Phase bin & Time (s)\\
& (from PN offset) \\
\hline
Pre-flare &  0  - 12482  \\
1st &  12482  - 14117 \\
2nd & 14117  - 15417 \\
Post-flare & 15417 - 19668 \\
Small bin 1 & 12845 - 13212 \\
Small bin 2 & 13212 - 13580 \\
Small bin 3 & 13580 - 13947 \\  
Small bin 4 & 13947 - 14315 \\
Small bin 5 & 14315 - 14682 \\
   \hline

\end{tabular}
Notes: Times for each of the phase bins used in the spectral fitting, where 1st and 2nd correspond to the first and second haves of the flare and the small bins lie within these (where statistics allow). All times are quoted from the PN offset of 92870774.5 s and correspond to the those shown in Figure 2.
\end{minipage} 

\end{center}
\end{table}

We select four high signal-to-noise ratio (SNR) phase bins covering the pre-flare, rise, fall and post-flare (see Table 1) and extract corresponding spectra. In addition we extract spectra from five consecutive phase bins (Table 1) sampling the highest count rates in the flare. By fitting simultaneously across all 9 spectra in the full 0.2-10~keV range (we note the calibration uncertainty between 0.2 and 0.3 keV although in this case our model is most sensitive to the changes between 0.3-1 keV) we obtain a higher time resolution picture of the evolution, which must be consistent with the highest SNR spectral models. In the spectral fitting we use only the PN data as the MOS data is of a poorer quality and applying the necessary multiple cross-normalisations is prohibitive. We note a likely instrumental feature in the PN spectrum at highest count rates not present in the MOS data, which we remove at the cost of 3 spectral bins (0.43-0.5 keV). 

\begin{figure}
\center
\includegraphics[width=3.5in, height=5in, angle=0]{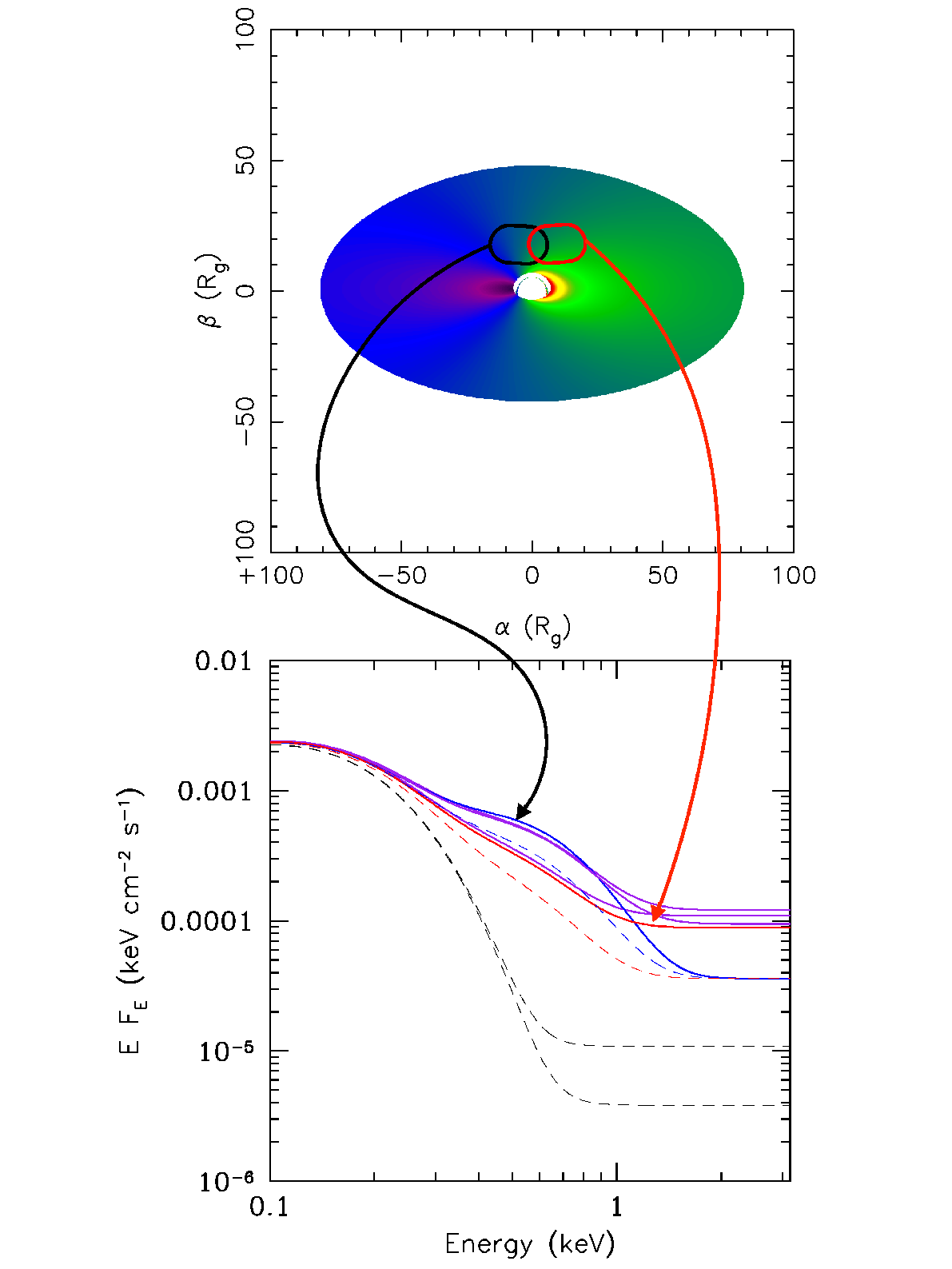}
\caption{Best-fitting phase-resolved spectral models.  Each phase-bin corresponds to a range of orbital positions in the wind, which leads to an evolution in spectral shape (bottom panel). We mark on the disc image the regions which contribute flux through the window during two such phase intervals, and point to their corresponding de-absorbed, best-fitting spectra. The black dashed models are the best fits to the `quiescent' emission (out of the flare) and the blue and red dashed spectra are the higher SNR spectra (1st bin and 2nd bin in Table 1). The solid line best-fitting models correspond to the shorter phase bins and are colour coded to represent the passage of the window from the blue-shifted to red-shifted side of the disc. As shown in Figure 4, this model fitting is tightly constraining (even given the relatively poor data quality available). The non-circular shape of the projected window shown here results from the window's passage over the duration of a phase-bin.}
\end{figure}

We construct a model in {\sc xspec} v12 (Arnaud 1996) comprising neutral absorption ({\sc tbabs}, at solar abundance Wilms, Allen \& McCray 2000), our window model ({\sc win}) and a power-law ({\sc pow}): {\sc tbabs*(win+pow)}. The model parameters for the window model are as follows: black hole dimensionless spin, inclination (limited to 40-70 degrees to provide inclinations that may intercept the wind but avoid the torus), mass accretion rate in units of the Eddington limit, black hole mass, window angular size ($\Delta_{\rm w}$ restricted to be $\le$ 50 degrees), window inclination ($\theta_{\rm w}$ measured from the black hole spin axis), disc wind outer radius ($d_{\rm w}$ restricted to be $\ge$ 80 $R_{\rm g}$), time of central passage, phase bin start time, phase bin end time and normalisation. 

We initially fix the absorption to be that from the best-fit to the mean spectrum (consistent with a Galactic column of 8$\times$10$^{19}$ cm$^{-2}$: Dickey \& Lockman 1990); this is a sensible precaution as the poorer quality data in the individual phase bins would incorrectly allow a greater span of column densities. We proceed to tie the spin and window properties (including normalisation) between phase bins and fix the mass at its best estimate of 8$\times$10$^{5}$ M$_{\odot}$ (Sun et al. 2013). As the emission is dominated by a thermal component (i.e. the Wien tail) and given the poor statistics above 1~keV, we fix the power-law index ($\Gamma$) to 2 to be consistent with the index seen in the highest signal-to-noise ratio data out of the `flare' and normalisations free to vary (except in the case of the 1st bin of Table 1 and Small bin 3 where we tie the normalisation to that of 2nd bin due to the poorer data quality).  We also place a hard limit on the maximum accretion rate of 10$\times$Eddington as only a handful of (radio-bright) QSOs have been seen to accrete at such levels for prolonged periods (Woo \& Urry 2002), whilst short-lived tidal disruption events at higher accretion rates are expected to be accompanied by powerful ejecta (e.g. Swift J1644+57: Levan et al. 2011), not reported for this AGN (which we note to be radio quiet). 

\begin{figure}
\begin{center}
\begin{tabular}{l}
 \epsfxsize=8cm \epsfbox{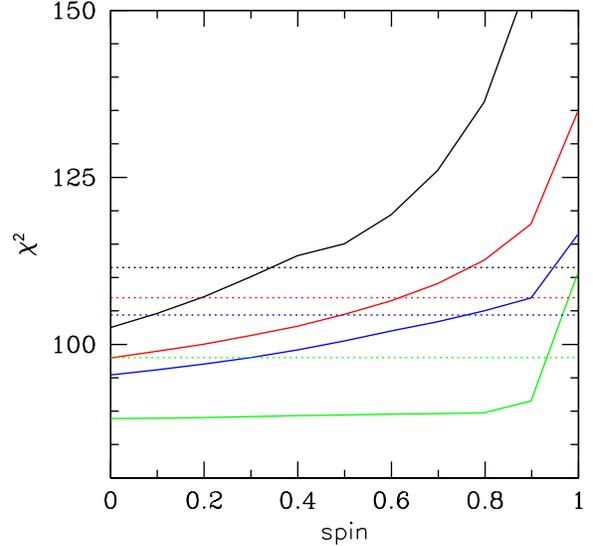}
\end{tabular}
\end{center}
\caption{We determine the $\chi^2$ space for the spin parameter by use of the {\sc steppar} command in {\sc xspec}. We perform this for the Shakura-Sunyaev profile disc (black), with free absorption (red), with the upper limit on the mass (green) and the free temperature profile disc (blue). The horizontal dotted line of the same colour corresponds to a $\Delta\chi^2$  = 9 (3-sigma significance). In all cases with the best-fit mass we show that the spin is $<$ 0.8 at $>$ 3-sigma and at the upper limit of the mass we rule out maximal spin at $>$ 3-sigma.}
\end{figure}

Assuming standard disc emission (a radial temperature profile of T$^{4} \propto$ R$^{-3}[1-\sqrt{R_{in}/R}]$) we find that the resulting fit quality to the several phase bins is good ($\chi^{2}$  = 102.5/90 d.o.f.s) with the best fitting model shown in Figure 3 and parameters with 90\% errors (from using the {\sc uncer} command) given in Table 2. We use the {\sc steppar} command to obtain the chi-squared contours for the spin parameter shown in Figure 4 (with 3-sigma confidence levels indicated), finding the spin (restricted to be prograde only for simplicity) to be $<$ 0.4 at $>$ 3-sigma confidence (with a best-fit at zero) and rule out maximal spin at $>$ 5-sigma. Our inferred inclination range (40-70 degrees) is fully consistent with the source resembling a Seyfert 1 in its optical spectrum (Sun et al. 2013) with a line-of-sight intercepting the wind yet still avoiding the dusty torus (Fischer et al. 2013).

To demonstrate the power of phase-resolved spectroscopy, we also fit the combined data over the `1st' and `2nd' phase bins (i.e. the flare) with a simple phenomenological model that incorporates the spin. For this we used the  {\sc xspec} model {\sc tbabs*kerrconv*(diskbb+pow)}; although more complex models exist, e.g. {\sc kerrbb} and {\sc bhspec}, these do not extend to AGN masses and the bespoke model, {\sc optxagnf} (Done et al. 2012) would require optical data to be relatively contemporaneous (the optical SDSS data was taken 7 years after the {\it XMM-Newton} observation - whilst we have a single OM datapoint this does not allow for tight constraints on the spectral shape). We find that the resulting spin value is unconstrained.

\begin{table}
\begin{center}
\begin{minipage}{82mm}
\bigskip
\caption{Best-fitting model parameters}
\begin{tabular}{lc|c|c|}
  \hline

{\sc tbabs*(win + pow)} &  T $\propto$ R$^{-3/4}$ & T $\propto$ R$^{-p}$\\
\hline
 Inclination (degrees) &  58$^{+12}_{-18}$ & 61$^{+4}_{-21}$\\
Mass accretion rate (Eddington) &  10 ($>$ 9) &  10 ($>$ 9)\\
$\Delta_{w}$ (degrees) & 10$^{+3}_{-2}$  & 15 $\pm$ 4\\
$\theta_{w}$ (degrees) & 45$^{+7}_{-17}$ &  48 $\pm$ 3*\\
$d_{w}$ ($R_{\rm g}$) & $<$ 82$^{\times}$ & $<$ 83$^{\times}$ \\
Emissivity (4$\times$p) & 3.0 &  2.6 $\pm$ 0.2\\
$\chi^2$/d.o.f & 103/90 &  95/90 \\  

   \hline

\end{tabular}
Notes: Best fitting model parameters for our phase binned spectroscopy. We initially assume a standard Shakura-Sunyaev emissivity profile (T $\propto$ R$^{-3/4}$) and then test a flatter profile, distorted due to advection (T $\propto$ R$^{-p}$). The parameters are those discussed in the text and the error bounds are quoted at 90\% significance. * denotes that the inclination was fixed in order to constrain the error range on $\theta_{w }$ (and to prevent the fit from moving to a unphysical parameter space) and should be treated as a lower limit. $^{\times}$ indicates that the lower limit of $d_{w}$ at 80 $R_{\rm g}$ has been met in determining the error bounds. The spin error contour plots are shown in Figure 4.
\end{minipage} 

\end{center}
\end{table}

\subsection{Testing the caveats}

The aforementioned spectral fitting relied on a series of caveats which we shall now explore:

\begin{itemize}{

\item{We fixed the absorption column to be the same throughout, i.e. 8$\times$10$^{19}$ cm$^{-2}$, consistent with Galactic column in the line-of-sight (Dickey \& Lockman 1990) and the best-fit value to the time-averaged spectra. Although a sensible precaution, we test whether this constraint affects the outcome, allowing all phase bins to have the same column but with Galactic column as the lower limit. We find a best-fitting spin of zero constrained to be $<$ 0.7 at $>$ 3-sigma (red line in Figure 4, with a column of $<$ 3$\times$10$^{20}$ cm$^{-2}$ at 3-sigma) and once again rule out maximal spin at $>$ 5-sigma.}\\

\item{We proceed to test whether setting the black hole mass to its maximum limit affects the result. Assuming a standard 0.5 dex error on the mass, we test at the upper limit of 2.4$\times$10$^{6}$ M$_{\odot}$ which would require a larger spin to reach higher temperatures. The best fit (with $d_{\rm w}$ altered to give an orbital period $\ge$ 18~ks at this mass), has a spin of zero and we rule out maximal spin at $>$ 3-sigma (see Figure 4, green line).}\\

\item{The putative range of accretion rate ($>$ 9.4$\times$Eddington) readily explains the presence of such an optically thick wind (Shakura \& Sunyaev 1973; Poutanen et al. 2007). However, it remains unclear at what radius in a super-critical AGN disc the temperature profile is likely to change (and indeed how flat a profile is created), however, we can test whether incorporating a flatter profile with T $\propto$ R$^{-p}$  (where 0.5 $< p <$ 0.75, Abramowicz et al. 1988), changes our findings. We fit this model to all phase bins and provide the interesting parameters and their 90\% error bounds in Table 2. We find the best-fitting spin to be zero, constrained to be $<$ 0.8 at $>$ 3-sigma (see Table 2 and Figure 4, blue line). We note that this is only a partial treatment of the change in disc structure at such high accretion rates, and the associated change in scale-height may lead to effects we have not fully accounted for, e.g. the effect of a non razor-thin disc in ray tracing and effect of self-shieilding.}\\

{\item We note that in BHBs around their Eddington limit, it is probable that the structure of the inner disc changes, perhaps due to the increased radiation pressure (Middleton et al. 2012), magnetic pressure support (Straub, Done \& Middleton 2013) or turbulence, leading to a Compton up-scattered component (e.g Ueda et al. 2009) equivalent to having a region of the disc with a higher colour temperature correction. Although it is unclear whether up-scattering plays a role in super-Eddington accretion rate systems (which should be wind dominated in the outer regions), should this be the case here we could have lower disc temperatures and therefore reach mass accretion rates more commonly seen in bright, radio-quiet AGN (observationally up to ~3$\times$Eddington: Woo \& Urry 2002).  However, as the evolution of the flare is {\it independent} of model assumptions, the result would remain unchanged with a best-fitting spin of zero.}
}

\end{itemize}

\section{Discussion \& Conclusion}

Predictions of hierarchical growth (Fanidakis et al. 2011) require low mass AGN to have maximal spin when the SMBH growth was constant/prolonged. However, determining the spin of low mass AGN is expected to prove difficult due to observational bias towards those at highest mass accretion rates, leading to weak signatures of reflection or the presence of optically thick winds distorting our view of the emission. Given the expectation of radiative-hydrodynamic instabilities in such winds, we predict that at favourable inclinations, our view of the inner disc will change as a function of the orbital timescale of the resulting inhomogeneities. This provides a new lever arm for determining the spin, where the phase-resolved spectrum is a function of spin and inclination (as well as mass accretion rate, window properties etc). This has been studied by Risaliti et al. (2011) from the perspective of changing reflection features in more massive AGN. Here we have created a model which incorporates ray tracing and full GR effects but instead recreates the changing shape of the accretion disc, expected to extend into the soft X-ray band for low mass AGN, allowing us to perform Doppler tomography directly.

RX J1301.9+2747 is an apparently unique AGN in terms of its long timescale appearance in the X-rays, showing short, $\sim$2~ks flares spaced by $\ge$18~ks but with a flat featureless lightcurve besides. We have argued that this unusual variability is most probably associated with a dynamical timescale and given its soft spectrum and UV excess we suggest that this may be the signature of a `window' in the wind passing over the inner disc. By fitting the changing spectrum over several phase bins simultaneously, we have shown that the spin is most likely very low even when accounting for a series of important caveats, notably that the accretion rate is expected to be very high indicating a deviation away from standard disc theory. 

Besides being the first spin estimate using the method of Doppler tomography, this would appear to be the lowest spin ever found for an AGN and the lowest mass AGN for which a spin has been determined (Reynolds 2013). Based on the expected spin distribution (Fanidakis et al. 2011) our result may point towards chaotic growth in agreement with the findings of Nayakshin, Power \& King (2012). Substantial numbers of low mass AGN have now been identified and, as most of these appear to be accreting at large fractions of Eddington (Greene \& Ho 2004) we expect the accompanying presence of optically thick winds. For an isotropic distribution of inclinations, although extremely rare (as demonstrated by the hitherto unique behaviour of this source), we should expect to apply this method to further sources and determine the spin distribution more robustly.

Although we have argued that our interpretation is the most likely in light of the timescales of variability, the spectra and the optical/UV excess, we cannot yet claim a {\it unique} identification. Indeed the flare lightcurve can be described by a fast rise exponential decay profile (FRED), which may result from an unknown accretion instability operating on (relatively) fast timescales, perhaps associated with the apparent high accretion rates. Further X-ray data will provide a view of many more flares, thereby helping to establish a unique interpretation for their origin in this source. In addition, our model fits (Table 2) would appear to suggest that the flares are periodic as $d_{w}$ is close to the value needed to produce an orbital timescale of $\ge$18~ks; further data will naturally test this model prediction.

We note that, irrespective of the nature of the flares in this particular source, the method itself is a powerful one and, as an exciting aside, we can invert the method to obtain the black hole spin for BHBs where the emission is regularly obscured (i.e. eclipsed) rather than revealed, this new approach may help solve the debate regarding whether `tapping' of the spin (Blandford \& Znajek 1977) leads to powerful jet events seen in both BHBs and AGN (Narayan \& McClintock 2012; Steiner, McClintock \& Narayan 2013; Russell et al. 2013; Middleton, Miller-Jones \& Fender 2014).

\section{Acknowledgements}
The authors thank the anonymous referee for their helpful comments. This work is based on observations obtained
with {\it XMM-Newton}, an ESA science mission with instruments and
contributions directly funded by ESA Member States and NASA.

\label{lastpage}

\appendix
\section{}

Here we describe the details of our Doppler Tomography model and the various caveats: 

\subsection{Coordinate system}

We use a right-handed coordinate system (shown in Figure A1) with the BH located at the origin and the BH spin axis coinciding with the z-axis. The vector $\mathbf{\hat{o}}$ (hat denotes a unit vector) pointing from the BH to the center of the observer's camera is in the x, z plane (i.e. it has no component in the y direction) and makes an angle $i$ with the z-axis. The observer is assumed to be far away such that parallax can be ignored; i.e. the vector pointing from any point near the BH to any point on the camera is assumed to be $\mathbf{\hat{o}}=(\sin i,0, \cos i )$. If the camera is some (large) distance $D$ from the BH, the vector pointing from the origin to any point on the camera is: $D\mathbf{\hat{o}} ~+~ \alpha\mathbf{\hat{j}} ~+~ \beta [\sin i ~\mathbf{\hat{k}}- \cos i ~ \mathbf{\hat{i}}]$, where $\alpha$ and $\beta$ are the impact parameters at infinity and $\mathbf{\hat{i}}$, $\mathbf{\hat{j}}$ and $\mathbf{\hat{k}}$ are the unit vectors in the x, y and z direction respectively. This means that any pixel on the camera is a distance $b=\sqrt{\alpha^{2}+\beta^{2}}$  from the center of the camera, consisting of a distance $\alpha$ in the horizontal direction and $\beta$ in the vertical direction. The angle between a vertical line on the camera and a line pointing from the centre of the camera to a pixel at $(\alpha, \beta)$ is thus given by $\tan\theta_{\rm b} = \beta / \alpha$. We assume that the window through which we can see the inner disc is a distance $d_{\rm w}$ from the BH. Here, we express all distances in units of $R_{\rm g}$.

\begin{figure*}
\center
\includegraphics[width=6in, angle=0]{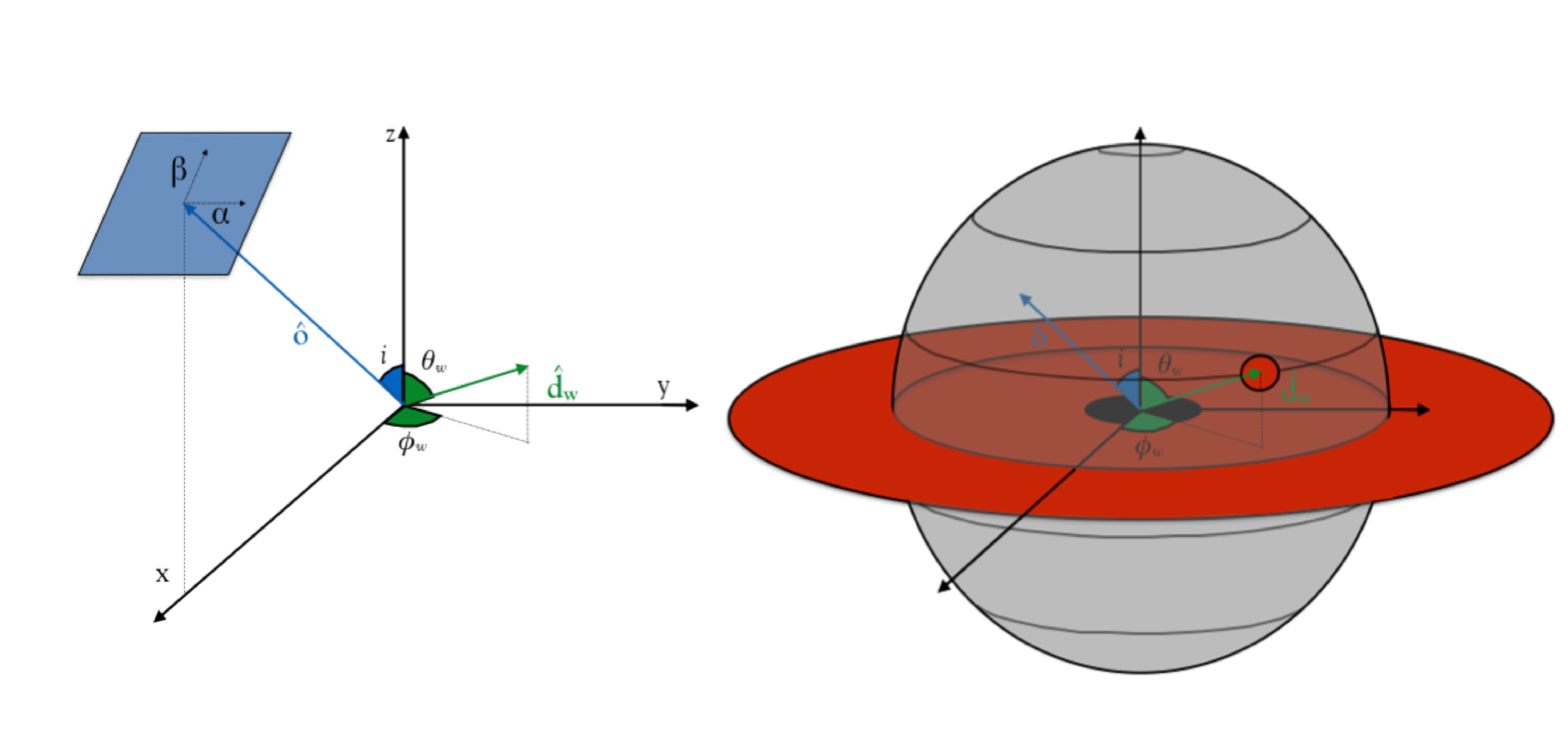}
\caption{Coordinate systems for our model. The left hand plot shows the coordinate system and arrangement of vectors connecting the observer  (blue square) to the black hole (at the centre of the space). For reference the same vectors are shown with a spherical shroud/wind, orbiting window and outer disc in the right hand plot. }
\end{figure*}

\subsection{Null geodesics}

According to general relativity (GR), if gravity is the only force and the only gravitating object is a point source (a BH is an excellent approximation to this scenario), the geodesic path of a photon escaping to infinity depends exclusively on the BH spin, the polar angle describing the position of the observer i, and the impact parameters at infinity, $\alpha$  and $\beta$. We use a grid of impact parameters with 200 logarithmic steps in $b$ and 200 linear steps in the angle $\theta_{b}$. This resolution gives a final spectrum in agreement with finer resolutions better than the 1\% level. For each combination of $\alpha$ and $\beta$, we use {\sc geokerr} to trace the geodesic path from this pixel back to infinity, or alternatively the BH horizon if that pixel falls within the BH shadow. We set {\sc geokerr} to return 100 points along each geodesic (in equal jumps of 1/$r$) and track backwards from the observer, calculating at each point the coordinate values (x, y, z), for that point.

The disc is assumed to be in the equatorial plane and extend between $r_{\rm in} = ISCO(a_{*})$ and $r_{\rm out} = 90 R_{\rm g}$ (the flux from this region dominates over that from further out), or $r_{\rm out} = d_{\rm w}$ if $d_{\rm w} < 90 R_{\rm g}$. If the z-coordinate of the geodesic changes sign between two points, the geodesic has crossed the BH equatorial plane. We use cubic interpolation between these two points to determine the $r$ and $\phi$ values for which the geodesic crossed the z=0 plane. If $r_{\rm out} > r > r_{\rm in}$, the photon has intercepted the disc. If $r < r_{\rm in}$, we carry on following the geodesic since it may loop back around and intercept the equatorial plane again. We note that by continuing to trace geodesics for $r < r_{\rm in}$, we have explicitly assumed that material between the ISCO and event horizon does not contribute to the spectrum.

\subsection{Observed spectrum for a given geodesic that intercepts the disc}

For geodesics that do intercept the disc, we calculate the rest frame temperature, $T(\rm r)$, and the energy dissipation per unit area, $\epsilon$(r), for the disc at the interception point according to standard disc theory:
\begin{eqnarray}
T(r) &=& f_{\rm CT} ~ 6.06 ~ {\rm keV} ~ \left[ \frac{ \dot{m}  r_{\rm in} }{M r^3} \left( 1 - \sqrt{ \frac{r_{\rm in}}{r} } \right ) \right]^{1/4} \\
\epsilon(r) &=& 3\left(\frac{r_{in}}{r}\right)^{3} \left(1-\sqrt{\frac{r_{in}}{r}}\right).
\end{eqnarray}
Here, $\dot{m}$ is the mass accretion rate in units of the Eddington limit and $M$ is the mass of the BH in units of solar masses. $f_{\rm CT}$ is a colour temperature correction ($f_{\rm CT}\approx$ 2.4 for AGN) which accounts for electron scattering in the disc atmosphere (Done et al. 2012). In addition, we assume the rest frame spectrum to be given by Plank's law of blackbody radiation, for the temperature $T(\rm r)$. We also test an alternative disk emissivity law $T(\rm r) \propto r^{-p}$, where $p$ is a model parameter. In this case, the normalisation constant for the flux and temperature as a function of $r$ is determined by equating the total luminosity for a disc extending from $r = r_{\rm in}\to\infty$ to that for the case with $p$ = 3/4.

Corrections are applied to the spectrum to account for the velocity of the disc material plus GR effects. We assume orbits with angular velocity $\omega(r) = (c/R_{\rm g})/(r^{3/2}+a)$ radians per second. For a metric with the line element $ds^{2} = {\bf g_{\mu\nu} dx^{\mu} dx^{\nu}}$, the resulting blue shift experienced by a photon emitted from the disk at r and incident on our camera at ($\alpha$, $\beta$) is given by:
\begin{equation}
\frac{E}{E'} = \frac{\sqrt{-{\bf g_{tt}}-2{\bf g_{t\phi}}-{\bf g_{\phi\phi}}\omega^{2}}}{1+ \omega\alpha \sin(i)},
\end{equation}
where E is the observed energy and $E'$ is the rest frame energy. We use the Kerr metric in the equatorial plane (see Dexter \& Agol 2009 for the full expression). For a derivation of the above equation, see Luminet (1979).

The specific flux contributed from the pixel at energy $E$ is then:
\begin{equation}
dF_{E} = \left(\frac{E}{E'}\right)^3 \epsilon(r) I'_{E'}(r,E') d\Omega,
\end{equation}
where d$\Omega$ is the solid angle subtended by the pixel, which is proportional to ($b~ db ~d\theta_{\rm b}$), and $I'_{\rm E'}(r,E')$ is the rest frame specific intensity at energy $E'$ (Veledina, Poutanen \&
Ingram 2013). The spectrum of radiation hitting each pixel is thus boosted/suppressed and shifted according to the point of the disc from which the photons emerged.

\subsection{Windowing}

We assume that a spherical shroud centred on the BH with radius $d_{\rm w}$ blocks the disc from view except for a window (Figure A1). The vector pointing from the BH to the centre of the window makes an angle $\theta_{\rm w}$ with the z-axis, and the opening angle of the window on the surface of the shroud is $\Delta_{\rm w}$. The window stays at a constant $\theta_{\rm w}$ and rotates with Keplerian velocity such that the azimuth of the window $\phi_{\rm w}$, defined from the x-axis, increases at a constant rate dependent on $d_{\rm w}$. The model parameter $t_{\rm 0}$ defines the time at which the centre of the window is at $\phi_{\rm w}$ = 0.

For each pixel (i.e. permutation of $\alpha$ and $\beta$), we work out for how long within the time interval $t_{\rm start}$ to $t_{\rm end}$ the disc is visible through the window. Since $d_{\rm w}$ is large, we can ignore light bending. If a given geodesic crosses the disc at $\mathbf{r} = (x, y, 0)$, the point on the shroud directly between the observer and $\mathbf{r}$ is defined by the vector $\mathbf{p} = \mathbf{r} + d_{\rm w}\mathbf{\hat{o}}$. If the angle between $\mathbf{p}$ and $\mathbf{d_{\rm w}}$ (the vector pointing from $\mathbf{r}$ to the center of the window) is less than $\Delta_{\rm w}$/2, we are looking through the window, otherwise our view is blocked. Defining $\alpha'$ = $\alpha$/d$_{\rm w}$, $\beta'$ = $\beta$/$d_{\rm w}$ and $\gamma'^{2} = (1-\alpha')^{2} - \beta'^2$, we can express $\mathbf{p}$ as
\begin{equation}
\mathbf{\hat{p}} = (\gamma' \sin(i) - \beta' \cos(i), \alpha', \gamma'  \cos(i) + \beta' \sin(i) ),
\end{equation}
and $\mathbf{\hat{d}_{\rm w}}$ as
\begin{equation}
\mathbf{\hat{d}_{w}} = ( \sin\theta_{w}  \cos\phi_{w}, \sin\theta_{w}  \sin\phi_{w}, \cos\theta_{w}).
\end{equation}

The window is visible if $\mathbf{\hat{d}_{w}} \cdot \mathbf{\hat{p}} < \cos(\Delta_{w}/2)$. So, we need to know, for what range of $\phi_{w}$ is this condition met? To answer this question, we must find the two critical $\phi_{w}$ angles, for which $\mathbf{\hat{d}_{w}} \cdot \mathbf{\hat{p}} = \cos(\Delta_{w}/2)$. These can be found by solving the equation $0 = A \cos\phi_{w} + B \sin\phi_{w} + C$, where
\begin{eqnarray}
A &=& \sin\theta_{w} [\gamma'  \sin(i)-\beta'  \cos(i)] \nonumber \\
B &=& \alpha' \sin\theta_{w} \nonumber \\
C &=& \cos\theta_{w}[\gamma'  \cos(i)+ \beta'  \sin(i)]-\cos(\Delta_{w}/2),
\end{eqnarray}
to get
\begin{equation}
\tan(\phi_{w}/2) = B \pm \sqrt{\frac{A^{2}+B^{2}-C^{2}}{A-C}}.
\end{equation}
This yields two solutions in the interval $-\pi$ to $\pi$: $\phi_{small} < \phi_{big}$; unless $(A^{2} + B^{2}) < C^{2}$, in which case the square root term is imaginary and there is no solution. If there is no solution, this indicates that for all values of $\phi_{w}$, the line-of-sight does not pass through the window. If $(A^{2} + B^{2}) = C^{2}$, this indicates there is one value of $\phi_{w}$ where the line-of-sight hits the outline of the window (i.e. the top or bottom). When there are two solutions, this indicates that the line-of-sight passes through the window for $\phi_{small} < \phi_{w} < \phi_{big}$. Thus, the line-of-sight passes through the window for a fraction ($\phi_{big} - \phi_{small}$)/2$\pi$ of the window's orbit. We may not be observing for the entire orbit though. The azimuth of the window relates to time as 
$\phi_w(t) = (t-t_0) \Omega_{k} $, where $\Omega_{k} = d_{w}^{-3/2} (c/R\d{\rm g})$ is the Keplerian angular velocity. The fraction of our observing time for which a particular line-of-sight passes through the window is thus
\begin{equation}
f = \frac{ {\rm MIN}[ \phi_{big}, (t_{end}-t_{0}) \Omega_{k} ] - {\rm MAX}[ \phi_{small} , (t_{start}-t_{0}) \Omega_{k} ] } { (t_{end} - t_ {start}) \Omega_{k} },
\end{equation}
or $f$ = 0 if the above expression yields a negative value (implying the window is never in the line-of-sight).

We calculate the observed spectrum from every pixel, $dF_E(E)$, and multiply this by the factor $f$, calculated for that particular pixel before summing to obtain the total spectrum observed from inside the shroud. As the spectrum is inevitably dominated by the emission from the hottest parts of the disc, the choice of window shape is relatively insensitive. However, given that an irregularity along the surface of the wind is likely in light of hydrodynamic instabilities (e.g. Rayleigh-Taylor) our choice of a circular aperture is both simple and appropriate.

\subsection{Outer disc}

We also see (considerably less) emission from outside the shroud, which contributes to the quiescent spectrum. Since this is far from the BH, we can safely ignore light bending and ignoring rotation agrees with a more careful treatment at the 1\% level, which is observationally indistinguishable at quiescent count rates. These assumptions allow a simple analytic treatment where we split the disc up into rings with radius $r$ and, for each ring, determine for what range of $\phi$ the line-of-sight is obstructed by the spherical shroud. If a point on the disc [described by $r\mathbf{\hat{r}}(\phi)$] is blocked by the shroud, we can follow the vector $\mathbf{\hat{o}}$(i) from that point for some distance $X$ until we are a distance $d_{w}$ from the BH (i.e. the border of the shroud). We can solve for $X$: if there is a real solution, the disc patch is blocked by the shroud, otherwise we see that patch of disc. $X$ can be found by solving the quadratic
\begin{equation}
0 = X^{2}+[2r \cos\phi  \sin(i) ]X+[r^{2}-d_{w}^{2}],
\end{equation}
and there is a real solution if the determinant is positive. Since our coordinate system is chosen such that we always see $\phi$ = 0, there is never a solution for $X$ at $\phi$ = 0 and we can exploit the symmetry around $\phi = \pi$. If $\phi \le \phi_{crit}$, we see that point on the disc and if $\phi_{crit} < \phi  \le \pi$, our view is obstructed by the shroud (note, for very large r, this means $\phi_{crit} > \pi$). Here $\phi_{crit}$ is the critical angle for which the determinant is equal to zero:
\begin{equation}
\cos\phi_{crit} =  \frac{\sqrt{1-(d_{w}/r)^{2} )}}{\sin(i)}.
\end{equation}

The observed fraction of a ring at r (i.e. not obstructed by the shroud) is then simply 
$f = |\phi_{crit}|/\pi$  (or $f$ = 1, depending on which is smaller). We calculate the flux for a grid of 100 rings with $r$ changing in equal logarithmic steps from $d_{w}$ to 9$d_{w}$ as
\begin{equation}
dF_{E'} = f \epsilon(r) I'_{E'} d\Omega
\end{equation}
where $\epsilon(r)$ is as before and, since we ignore light bending, the solid angle is simply
\begin{equation}
d\Omega  \propto 2\pi ~ r ~ dr ~ \cos(i),
\end{equation}
and, since we ignore rotation, $dF_E = dF_{E'}$. We then sum over all rings and add this contribution to the spectrum from inside the shroud.

\end{document}